\documentclass[a4paper, notoc]{JHEP3}
\usepackage{amsfonts}
\usepackage{amssymb}
\usepackage{epsfig}
\usepackage{cite}

\newcommand{\conho}{\tilde}
\newcommand{\conhoya}{\widetilde}
\newcommand{\fn}{\footnote}
\newcommand{\beqa}{\begin{equation}}
\newcommand{\eeqa}{\end{equation}}
\newcommand{\p}{\partial}
\newcommand{\oh}{{1\over 2}}
\def\beqa{\begin{eqnarray}}
\def\eeqa{\end{eqnarray}}

\title{Holography, The Second Law and a $\conhoya{C}$-Function
in Higher Curvature Gravity}
\author{Daniel Cremades$^*$, Ernesto Lozano-Tellechea$^{**}$\\ *\ DAMTP,
  Centre for Mathematical
Sciences,\\
  Wilberforce Road, Cambridge, CB3 0WA, UK.\\ 
**\ Department of Particle Physics, Weizmann Institute of Science\\
Rehovot 76100, Israel\\ \vspace{2mm}
  E-mail:    
  \email{d.cremades@damtp.cam.ac.uk} ,  \email{ernesto.lozano@weizmann.ac.il}}
 
\abstract{We analyze the Second Law of black hole mechanics and the
generalization of the holographic bound for general theories of
gravity. We argue that both the possibility of defining a holographic
bound and the existence of a Second Law seem to imply each other via
the existence of a certain ``c-function'' (i.e. a never-decreasing
function along outgoing null geodesic flow). We are able to define
such a ``c-function'', that we call $\conhoya{C}$, for general
theories of gravity. It has the nontrivial property of being well
defined on general spacelike surfaces, rather than just on a spatial
cross-section of a black hole horizon. 
We argue that $\conhoya{C}$ is a suitable generalization of the concept of ``area" in any extension of the holographic bound 
for general theories of gravity.
Such a function is provided by
an algorithm which is similar (although not identical) to that used by
Iyer and Wald to define the entropy of a dynamical black hole. In a
class of higher curvature gravity theories that we analyze in detail,
we are able to prove the monotonicity of $\conhoya{C}$ if several
physical requirements are satisfied. Apart from the usual ones, these
include the cancellation of ghosts in the spectrum of the
gravitational Lagrangian. Finally, we point out that our
$\conhoya{C}$-function, when evaluated on a black hole horizon,
constitutes by itself an alternative candidate for defining the
entropy of a dynamical black hole.}

\preprint{DAMTP-2006-66\\
WIS/12/06-AUG-DPP\\ hep-tr/0608174}

\begin{document}

\tableofcontents

\section{Introduction}

It is known that, in a general theory of gravity including higher
derivative couplings, the entropy of a stationary black hole is no
longer given by a quarter of the area of the event horizon. It was
shown in~\cite{Wald:1993nt,Iyer:1994ys} that the definition of entropy
which obeys the First Law of black hole mechanics is, instead, given by the
integral of a particular local quantity on a spatial cross-section
$\Sigma$ of the event horizon. In a theory which does not depend on
covariant derivatives of the Riemann tensor, the expression for the
entropy is:
\begin{equation}
   \label{WaldEntropy}
   S_{\rm BH} =
   -2\pi\int_{\Sigma}\frac{\partial L}{\partial R_{abcd}}\,
   \epsilon_{ab}\, \epsilon_{cd}\,
   \sqrt{h}\, d\Omega\, .
\end{equation}
where the Lagrangian of the theory is ${\cal L}=\sqrt{-g}\, L$
(i.e. $L$ is a scalar), partial derivatives with respect to the
Riemann tensor have to be taken as if it was a quantity independent of
the metric\footnote{\label{RiemannDependence}It was shown
in~\cite{Iyer:1994ys} that, in any covariant theory, the dependence of
$L$ on the derivatives of the metric can always be written as a
dependence on the Riemann tensor and its covariant derivatives only,
i.e. $L=L(g_{ab}, R_{abcd}, \cdots)$. The derivative
in~(\ref{WaldEntropy}) means the derivative of $L$ when rewritten in
such a way.}, $\epsilon_{ab}$ denotes the binormal to the horizon
cross-section (i.e. the volume element in the 2-space perpendicular to
it) normalized so that $\epsilon_{ab}\epsilon^{ab}=-2$, and
$\sqrt{h}\, d\Omega$ is the volume element induced on $\Sigma$. The
overall normalization is for units $G_N=1$. In General Relativity the
above expression reduces to one quarter of the area of $\Sigma$ but,
in more general theories, it differs from it. \\

This generalization of the expression for BH entropy poses two
important questions:
\begin{itemize}
  \item[{\bf 1}] Does this expression for $S_{\rm BH}$ obey a Second
  Law of black hole mechanics?
  \item[{\bf 2}] How should we generalize the notion of ``area of the
  boundary'' of an arbitrary phy\-sical system in establishing any
  holographic bound in a general theory of gravity?
\end{itemize}

Question~{\bf 1} above was considered in~\cite{Jacobson:1995uq} for a
class of theories where the gravitational Lagrangian is a function of
the scalar curvature, i.e. $L=L(R)$. It was shown that, if certain
conditions of ``positivity of energy'' (together with Cosmic
Censorship and assuming asymptotic flatness, as in the GR case) are
satisfied, then a Second Law holds for $S_{\rm BH}$ in the $L(R)$
theories whenever the equations of motion are obeyed. In general,
proving a Second Law is a difficult task, since it is not clear which
physical conditions have to be obeyed in a general theory of gravity
for the Second Law to hold. Also, imposing the equations of motion is
straightforward in principle, but technically difficult in
practice. \\

On the other hand, Question~{\bf 2} arises because, as it is well
known, the upper limit on the degrees of freedom of a system being set
by the area of its boundary comes from black hole physics. The
original argument of 't~Hooft~\cite{'tHooft:1993gx} uses the fact that
any physical system confined in volume $V$ has to collapse to form a
black hole long before its entropy exceeds the area of the boundary of
$V$ in Planck units. Adding entropy to the resulting black hole would
just increase its size. Therefore, the maximal entropy that can be
physically realized in a given volume is that of a black hole of the
same size. However, when we go further and use this fact to establish
a universal holographic bound, we are concerned with the intrinsic
properties of the boundary of an arbitrary spacetime region. I.e. we
do not restrict ourselves to the case in which the physical system is
a black hole, nor to the case where the boundary of the system is an
event horizon. Consequently, when the entropy of a black hole is no
longer given by its area, Question~{\bf 2} arises naturally.

The first natural guess to answer Question~{\bf 2} would be to use
equation~(\ref{WaldEntropy}) but, instead of restricting ourselves to
the cross-section of the event horizon of a black hole, simply perform
the integral over the surface of interest (the boundary of the
spacetime region considered). However, this is not satisfactory since
(as discussed originally in~\cite{Wald:1993nt} and further
in~\cite{Jacobson:1993vj} --- see also~\cite{Iyer:1994ys}) the local
quantity that one integrates in~(\ref{WaldEntropy}) is only well
defined when evaluated on an spatial cross-section of a Killing
horizon, but it is ambiguous when evaluated on a general spacelike
surface.

Below we will show that a generalized concept of ``area of the
boundary'' (which reduces to~(\ref{WaldEntropy}) in the case of a
black hole horizon) exists and it is well defined for general,
spacelike, codimension two surfaces, irrespective of them being cross
sections of a Killing horizon or not. Our definition will be very
close to the proposal in~\cite{Iyer:1994ys} for the entropy of a
dynamical (i.e. non-stationary) black hole, although it will differ
slightly from it. \\

In this paper we would like to emphasize the fact that Questions~{\bf
1} and~{\bf 2} are related to a third one, namely:
\begin{itemize}
  \item[{\bf 3}] Can we define a ``holographic c-function'' in general
  theories of gravity?
\end{itemize}

Let us stress that we will use the name ``c-function'' throughout the
paper just by analogy, and that by this we do not necessarily mean the
true holographic dual of a field theory c-function. By a
``c-function'' we just want to refer here to a non-decreasing function
along ``outgoing'' null geodesic flow (i.e., for flat/$AdS$/$dS$
asymptotics, a non-decreasing function of the radial variable). We
will see that monotonicity of such a ``c-function'' is the behaviour
required on physical grounds to address the Second Law. In particular,
we will show below that, in $AdS$, this function is a strictly
increasing one, and not a constant as one would expect for a true
field-theory c-function in a CFT\footnote{We will elaborate a bit more
on this in Section~\ref{sec:AdS}.}. Nevertheless, we shall keep
sometimes the name ``c-function'' for it, but we prefer to denote it
by $\conhoya{C}$. In the case of Einstein gravity coupled to arbitrary
matter fields obeying the null energy condition, the existence of such
a function in static backgrounds was proved recently
in~\cite{Goldstein:2005rr}. Earlier attempts to define a (``true'')
holographic c-function include~\cite{Sahakian:1999bd}\fn{Considerations on holography in higher curvature 
gravity from a different perspective have recently appeared in  \cite{Mukhopadhyay:2006vu}.}.\\

Roughly, the basic relation between Questions~{\bf 1}, {\bf 2} and
{\bf 3} can be put like this: assume that $\conhoya{C}$ exists and
that it is well defined on arbitrary spacelike surfaces which need not
be the horizon of a black hole. Then, if $\conhoya{C}$ can be shown to
equal the black hole entropy when evaluated on a spatial cross-section
of a black hole horizon, then it clearly constitutes a natural
candidate for the answer of Question~{\bf 2}. Assume further that
$\conhoya{C}$ is non-decreasing along a congruence of outgoing radial
null geodesics. Then this means that it cannot decrease along the
affine parameter of the null geodesic generators of the horizon, which
in turn implies the Second Law. Along the paper we will elaborate more
carefully on all these statements. \\

Our considerations are completely general and, in particular, our
definition of the $\conhoya{C}$-function should be applicable in
general theories of gravity. However, as an application of our
proposal, in this paper we restrict ourselves to a particular class of
theories. These are the theories with a Lagrangian ${\cal L} =
\sqrt{-g}\, L$ given by:
\begin{equation}
  \label{LagrangianRPQ}
  L = L_{g}(R,P,Q) + L_{m}(g_{ab}, \psi, \partial\psi)\,
\end{equation}
where $R$ denotes the Ricci scalar, $P=R_{ab}\, R^{ab}$,
$Q=R_{abcd}R^{abcd}$, and $\psi$ denotes any matter fields\footnote{We
are assuming that $L_{m}$ that matter fields do not couple explicitly
to the curvature, and that further couplings do not arise from terms
of the form~$\nabla\psi$ (see footnote~\ref{RiemannDependence}). This
is always the case of any theory with arbitrary couplings to scalar,
axion and gauge fields.}. The consistency of these theories has been
studied recently in~\cite{Navarro:2005da}. We will show below that,
for these theories, if the matter fields obey the null energy
condition and, further, if the theory is ghost-free, a
$\conhoya{C}$-function exists which satisfies all the above
requirements. Throughout the paper we will restrict ourselves to four
dimensions, but our results should generalize in a straightforward way
to any number of spacetime dimensions. \\

Finally, let us mention that our definition of ${\conhoya{C}}$
constitutes, by itself, an alternative candidate to the proposal
in~\cite{Iyer:1994ys} for the definition of the entropy of a dynamical
black hole. We will comment a bit on this in the Conclusions. \\

\section{Holographic c-functions, Raychaudhuri equation and the Second Law}

In order to fix ideas, let us start by reviewing the result
of~\cite{Goldstein:2005rr} as well as some properties of the
Raychaudhuri equation and the Second Law of black hole dynamics in GR.

\subsection{A c-function in General Relativity}

The authors of~\cite{Goldstein:2005rr} showed that, in four
dimensional Einstein gravity coupled to any matter fields subject to
the null energy condition, any spherically symmetric, asymptotically
flat spacetime:
\begin{equation}
  \label{SphSymmST}
  ds^2 = -a(r)\, dt^2 + a^{-1}(r)\, dr^2 + b(r)\, d\Omega^2\, ,
\end{equation}
admits a ``c-function'', $\conhoya{C}$, which is given by the area
${\cal A}$ of a transverse sphere at radius $r$:
\begin{equation}
  \label{C}
  \conhoya{C} \equiv \frac{1}{4}\, {\cal A}(r) = \pi b(r)\, .
\end{equation}
It was shown in~\cite{Goldstein:2005rr} that the equations of motion
imply that $\conhoya{C}$ is a never-decreasing function of the radial
variable. Note that, in a black hole spacetime, $\conhoya{C}$ at the
horizon equals the black hole entropy.\\

\subsection{Relation to Raychaudhuri equation}
\label{sec:RelationToRaych}

As already noted in~\cite{Goldstein:2005rr}, the monotonicity of
$\conhoya{C}$ can be understood as a simple consequence of the
Raychaudhuri equation. Let us briefly review this. \\

Consider an arbitrary spacetime (static or otherwise) with metric
$g_{ab}$, and let $p$ be a point of this spacetime. Consider a null
geodesic congruence in a vicinity of $p$ and, in particular, the
geodesic of the congruence passing through that point. Let $k\equiv
k^a\partial_a=d/d\lambda$ be the corresponding tangent vector,
$\lambda$ being the affine parameter of the congruence. Define another
null vector on the tangent space of $p$, $n\equiv
n^a\partial_a=d/d\sigma$, satisfying $k^a n_a=-1$ and $k^a\nabla_a
n^b=0$. Consider finally two linearly independent spacelike vectors
$\eta_{(i)}^a\partial_a=d/dx^i$, $i=1,2$, orthogonal to $k$ and
$n$. The corresponding dual forms define a differential volume form in
the space orthogonal to $k^a$ and $n^a$ given by:
\beqa
\label{dV}
dV=\sqrt{h}~dx^1\wedge dx^2\, ,
\eeqa
with
\beqa
h_{ij}=\eta^c_{(i)}\eta^d_{(j)}\, g_{cd}\, .
\eeqa
Raychaudhuri equation tells us about the local behaviour of the
expansion ${\vartheta}$ of the congruence. The expansion is given by:
\begin{equation}
  \label{Expansion}
  \vartheta  = \frac{d\sqrt{h}/d\lambda}{\sqrt{h}} =
  \frac{d \log {\cal A}}{d\lambda}\, ,
\end{equation}
where ${\cal A}$ is the transverse area spanned by the congruence
(i.e. the magnitude of an area element of the ``wavefront''). From
this expression one demonstrates that:
\beqa {d\vartheta\over d\lambda}=-\oh \vartheta^2 -
\sigma^{ab}\sigma_{ab}-R_{ab}k^a k^b,
\label{rayc}\eeqa
where $\sigma^{ab}$ is the shear of the congruence. This is
Raychaudhuri equation. Note that this equation relies only on the
geometry of the spacetime, and hence is independent of the dynamics of
the theory. From this expression, using the fact that, in General
Relativity, $T_{ab}k^ak^b=R_{ab}k^ak^b$, and assuming the null energy
condition ($T_{ab}\zeta^a\zeta^b\geq 0$ for every null vector
$\zeta^a$), one gets the inequality:
\beqa
\label{RaySimp}
{d\vartheta\over d\lambda} \leq - \frac{1}{2}\vartheta^2 \leq 0\, .
\eeqa

Assume now asymptotic flatness. This means, in particular, that we
have a well defined ``radial coordinate'' $r$. If we take the
congruence to be ``outgoing'' (i.e. $dr/d\lambda>0$) then we have, at
infinity, ${\cal A}\sim r^2$. Therefore we get:
\begin{equation}
  \vartheta \rightarrow 0^+\,
\end{equation}
asymptotically. Assume now that $\vartheta$ cannot diverge. This,
together with the fact that $d\vartheta/d\lambda\leq 0$, means that:
\begin{equation}
  \label{ThetaIsPositive}
  \vartheta\geq 0\, ,
\end{equation}
for all $\lambda$, and therefore ${\cal A}$ can never be decreasing
along the outgoing null geodesic flow. \\

Concerning the behaviour of $\vartheta$, we note that it can only
diverge at a naked singularity (forbidden if one assumes Cosmic
Censorship) or at a caustic. However, it is easy to prove from
equation~(\ref{RaySimp}) that $\vartheta$ cannot diverge at any finite
affine parameter $\lambda$ if, initially, we have
$\vartheta(\lambda=0)>0$. In such a case this implies the existence of
a ``c-function'' $\conhoya{C}\sim {\cal A}$. Conversely, if a
``c-function'' $\conhoya{C}$ exists, then $\vartheta$ is never
negative along outgoing null geodesic flow. We will next see that, in
such a case, this immediately implies the Second Law.\\

Consider now the particular case of a static
spacetime~(\ref{SphSymmST}). In such a case ${\cal A}$
in~(\ref{Expansion}) is precisely the area of transverse spheres, as
in eq.~(\ref{C}). In terms of the radial coordinate of the
metric~(\ref{SphSymmST}) this just means that the ``c-function''
$\conhoya{C}$ defined in~(\ref{C}) is a non-decreasing function of
$r$~\cite{Goldstein:2005rr}\footnote{Previous attempts to find
holographic c-functions have also been based on the Raychaudhuri
equation~\cite{Sahakian:1999bd}.}.

\subsection{Relation to the Second Law}
\label{sec:RelTo2Law}

Consider now a black hole spacetime. Establishing the Second Law of
black hole mechanics amounts to prove that:
\begin{equation}
  \label{2ndLawinGR}
  \frac{dS_{\rm BH}}{d\lambda} =
  \int_\Sigma \vartheta\, \sqrt{h}\, d\Omega \geq 0\, ,
\end{equation}
where, in the expression above, $\Sigma$ is a cross-section of the
event horizon, $\sqrt{h}$ is an area element of it, and $\vartheta$ is
defined along the outgoing null geodesic congruence orthogonal to
$\Sigma$ ($k^a$ in~(\ref{rayc}) becomes here the null vector field
tangent to the horizon generators). Therefore, to
prove~(\ref{2ndLawinGR}) it is enough to prove that $\theta\geq 0$ at
every point along the (future directed) generators of the (future)
event horizon \cite{Hawking_2nd_law}.

The latter requirement ($\vartheta\geq 0$ along outgoing null geodesic
flow and, in particular, also along the horizon generators) is
precisely what we have just proved in the previous paragraph,
eq.~(\ref{ThetaIsPositive}). Moreover, it is well known that
$\vartheta<0$ at any point along the null horizon generators violates
Cosmic Censorship~\cite{Hawking_2nd_law}. Recall that the requirements
to prove that $\vartheta>0$ this were the EOMs to be satisfied, the
null energy condition, asymptotic flatness and Cosmic Censorship.

However, as we have seen, this is equivalent to the statement of the
$\conhoya{C}$-function defined in~(\ref{C}) being monotonic. Although
in~\cite{Goldstein:2005rr} the function $\conhoya{C}$ was originally
defined just for static spacetimes, this symmetry property was not
used at all in the proof of it being monotonic. The only thing that we
need is the function $\conhoya{C}$ to be defined along null geodesic
flow (the latter being spherically symmetric or otherwise). To relate
monotonicity of $\conhoya{C}$ to the Second Law, the only additional
requirement is that, if $\conhoya{C}$ is evaluated on a black hole
horizon, it should equal its entropy\footnote{Heuristically, note that
the fact that $\conhoya{C}={\cal A}/4$ is a non-decreasing function of
$r$ fits nicely with the holographic principle (``the total number of
degrees of freedom in a region is bounded by the area of its
boundary'') and the meaning of the c-theorem in field theory (``the
number of degrees of freedom cannot decrease along the RG
flow'').}.

\section{$L_g(R)$ theories}
\label{LRTHeories}

All these results reviewed above are valid in General Relativity. Next
we wish to consider the simplest class of Lagrangians of the
form~(\ref{LagrangianRPQ}) in which the gravitational Lagrangian is
only a function of the scalar curvature, $L_g=L_g(R)$. The validity of
the Second Law for these theories was considered
in~\cite{Jacobson:1995uq}. Here we want to review their result,
discuss the Second Law, and finally obtain a consistency condition
that will be used later in the general case of arbitrary Lagrangians
of the form~(\ref{LagrangianRPQ}).

\subsection{``Generalized'' Raychaudhuri equation}

Consider an arbitrary spacetime and a null geodesic congruence defined
as in Section~\ref{sec:RelationToRaych}. We define now the
function:
\begin{equation}
\label{ConhoInLR}
\conhoya{C} = -2\pi \frac{\p L_g}{\p R_{abcd}}
              \epsilon_{ab}\epsilon_{cd}\, \sqrt{h}
            =  4\pi L_R \sqrt{h}\, ,
\end{equation}
where $\epsilon_{ab}=k_an_b-k_bn_a$ stands for the binormal of the
surface defined by $dV$ in eq.~(\ref{dV}), and we define $L_R$ as $L_R
\equiv \partial L_g/\partial R$. In analogy to the GR case, we now
define:
\beqa \conhoya{\vartheta}\equiv {d\log\conhoya{C}\over
d\lambda}=\vartheta+{1\over L_R} k^a\nabla_a L_R \eeqa
(where $\vartheta$ is the expansion defined as in
eq.~(\ref{Expansion})). Hence:
\beqa \label{dvarthetadlambda}
{d\conhoya{\vartheta}\over d\lambda} =-\oh \vartheta^2 -
\sigma^{ab}\sigma_{ab}-R_{ab}k^a k^b -{1\over L_R^2}\left( k^b\nabla_b
L_R\right)^2+{1\over L_R} k^a k^b\nabla_a\nabla_b L_R, \eeqa
where we have used the Raychaudhuri equation. On the other hand, the
Einstein equations for a Lagrangian of the form~(\ref{LagrangianRPQ})
with $L_g=L_g(R)$ are given by:
\begin{equation}
  \label{EOMsR}
  L_R\, R_{ab} +
  \left(\nabla^2 L_R
       -\frac{1}{2} L_g \right) g_{ab} -
  \nabla_{a}\nabla_{b} L_R =
  T_{ab}\, ,
\end{equation}
where:
\begin{equation}
  T_{ab} = \frac{1}{2} L_m\, g_{ab} -
           \frac{\partial L_m}{\partial g^{ab}}\, .
\end{equation}
Since $k^a$ is null, we get, from~(\ref{EOMsR}):
\beqa T_{ab}k^ak^b=L_R\, R_{ab}\, k^a k^b - k^a k^b\nabla_a
\nabla_b L_R\, . \eeqa
Finally, using the above equation in~(\ref{dvarthetadlambda}), we
have:
\beqa \label{GRE} {d\conho{\vartheta}\over d\lambda} =-\oh
\vartheta^2 - \sigma^{ab}\sigma_{ab} -{1\over L_R^2}\left(
k^b\nabla_b L_R\right)^2-{1\over L_R} T_{ab} k^a k^b\, . \eeqa
This implies that:
\begin{equation}
   \label{GeneralizedInequality}
   \frac{d\conhoya\vartheta}{d\lambda} \leq 0
\end{equation}
whenever the null energy condition holds {\em if}, additionally:
\begin{equation}
  \label{Condition}
  L_R > 0\, .
\end{equation}
Equation~(\ref{GeneralizedInequality}) is the analog to the
inequality~(\ref{RaySimp}) obtained in GR from the Raychaudhuri
equation for a congruence of null geodesics, in which the area ${\cal
A}$ swept by the congruence has been replaced here
by~$\conhoya{C}$. \\

The ``generalized Raychaudhuri equation''~(\ref{GRE}) implies that
$d\conhoya{\vartheta}/d\lambda\leq 0$ if the condition $L_R>0$ is
met\footnote{Note that this is linked to the matter satisfying the
null/``antinull'' energy condition.}. Such a condition is dependent
both on the theory and on the specific background considered. However,
suppose that $L_R<0$ for a specific background whose Ricci scalar we
denote by $R_0$. Then, the weak field expansion of the metric around
such a background would be the one obtained from the Lagrangian:
\begin{equation}
  \label{EffectiveGN}
  L_g = L_g(R_0) +
        \left(\frac{\partial L_g}{\partial R}\right)_{R=R_0}
        (R-R_0) + \cdots\, ,
\end{equation}
and therefore we would get a negative gravitational coupling in the
spacetime regions where~(\ref{Condition}) does not
hold\footnote{\label{FNOnEffectiveGN}Remember that the Newton constant
$G_N$ is defined by the coefficient of the term $\sim(\partial h)^2$
in a weak field expansion of the metric, $\delta g_{ab}\sim
h_{ab}$.}. Note that this condition (positivity of the effective
$G_N$) is a natural generalization of the null energy condition for
this class of theories, since the n.e.c. can be rephrased as the
requirement of gravity being universally attractive. \\

These results (with focus on the Second Law, which we now turn to
discuss) were derived in~\cite{Jacobson:1995uq}, and the same
consistency condition~(\ref{Condition}) was obtained
there. In~\cite{Jacobson:1995uq} the condition $L_R>0$ was derived by
going to an auxiliary theory (dynamically equivalent to the
$L_g(R)$-theory) described by Einstein gravity coupled to an scalar
field.

\subsection{A ``c-theorem'' and the Second Law}

Eq.~(\ref{GRE}) means that, when $L_R>0$:
\begin{equation}
  \label{FirstDeriv}
  \conhoya{\vartheta} = \vartheta + \frac{d L_R/d\lambda}{L_R}
\end{equation}
is a decreasing function. Consider now an asymptotically flat
background (with radial coordinate $r$) and, as in
Section~\ref{sec:RelationToRaych}, a congruence of ``outgoing'' null
geodesics (i.e. $dr/d\lambda~>~0$). In such a case we have,
asymptotically:
\begin{equation}
  R \sim 1/r^3\, , \ \ \ \
  \vartheta \sim 1/r > 0\, ,
\end{equation}
since, in asymptotically flat spacetime of dimension $D$, all the
components of the Riemann tensor vanish at least like $\sim
1/r^{D-1}$. (We have that $\vartheta \sim 1/r > 0$ asymptotically by
the same argument used in Section~\ref{sec:RelationToRaych}.)  If the
gravitational interaction is of the form $L_g = R + {\cal O}(R^2) +
\cdots$ we find that:
\begin{equation}
  \conhoya{\vartheta} \rightarrow 0^+\, ,
\end{equation}
asymptotically.  Assuming again Cosmic Censorship (and by an argument analogous to that used in Section 2.2) this implies,
together with the fact of $\conhoya{\vartheta}$ being
monotonically decreasing with $r$, that:
\begin{equation}
  \label{MeCagoEnCristo}
  \conhoya{\vartheta}\geq 0
\end{equation}
for all values of $r$, meaning that the function $\conhoya{C}$ defined
in eq.~(\ref{ConhoInLR}) is a never decreasing one along the null
geodesic congruence. Let us stress again that, conversely, if one
proves monotonicity of $\conhoya{C}$, then the
condition~(\ref{MeCagoEnCristo}) follows immediately\footnote{We insist
on this since it is~(\ref{MeCagoEnCristo}) what implies the Second
Law.}.\\

Note that, by construction, in the case of a black hole spacetime,
$\conhoya{C}$ equals the black hole entropy~(\ref{WaldEntropy}) when
evaluated at the horizon. The analogous of eq.~(\ref{2ndLawinGR}) for
this class of theories is therefore~\cite{Jacobson:1995uq}:
\begin{equation}
  \label{2ndLawinLR}
  \frac{dS_{\rm BH}}{d\lambda} = \int_\Sigma \conhoya{\vartheta}\,
  \sqrt{h}\, d\Omega \geq 0\, .
\end{equation}
The relation the ``c-theorem'' just proved to the Second Law of black
hole mechanics is exactly analogous to the proof sketched in
Section~\ref{sec:RelTo2Law} if we replace $\vartheta$ by
$\conhoya{\vartheta}$. The fact that, also in this case,
$\conhoya{\vartheta}<0$ at the horizon generators also violates Cosmic
Censorship was proved in~\cite{Jacobson:1995uq}.

\section{A $\conhoya{C}$-function in higher curvature theories}
\label{sec:ConstructionOfTheConho}

In the last sections we have seen that it is possible to define a
function $\conhoya{C}$ in GR and in theories with gravitational
Lagrangian $L_g(R)$ which has the following properties:
\begin{itemize}
  \item[a)] It can be evaluated in any arbitrary spacelike surface.
  \item[b)] When evaluated at the event horizon of a black hole it equals
  its entropy.
  \item[c)] If certain physical conditions and certain boundary
  conditions are satisfied, then $\conhoya{C}$ is a non-decreasing
  function along outgoing null geodesic flow.
\end{itemize}
Given a $\conhoya{C}$-function with these properties, the following
consequences are implied:
\begin{itemize}
  \item{a) and~b) make of $\conhoya{C}$ a natural candidate to
  generalize the notion of ``area'' in any generalization of the
  holographic bound in these higher curvature theories.}
  \item{b) and~c) imply the Second Law of black hole mechanics in
  these theories.}
\end{itemize}
 
These conclusions explicitly answer Questions {\bf 1} and {\bf 2} in
the Introduction. At this point, however, we have to stress the
following issue: in more general theories of gravity, a
$\conhoya{C}$-function defined as in (\ref{ConhoInLR}), i.e.:
\begin{equation}
  \label{kkdevaka}
  \conhoya{C}\sim \frac{\p L}
  {\p R_{abcd}}\epsilon_{ab}\epsilon_{cd}\sqrt{h}
\end{equation}
cannot be well defined in general. The ambiguity arises from the fact
that one can always add a total divergence or a topological
term\footnote{\label{Ambiguities}By this we mean terms like, for
instance, a Gauss-Bonnet density in four dimensions. Such term just
adds a topological constant to the action (the Euler number of the
manifold) and thus it does not affect the equations of motion.} to the
Lagrangian. This addition leaves the dynamics of the theory unchanged
but, on the other hand, if such terms depend on the curvature, they will
clearly change the expression of $\conhoya{C}$ if defined as
above\footnote{\label{Footnote4}For simplicity, in this paper we
restrict ourselves to the case in which $L$ does not depend on
derivatives of the Riemann tensor. Therefore, in four dimensions, the
only cases we could be concerned with are the ambiguities arising from
the addition to the Lagrangian of a Gauss-Bonnet density (first Euler
class) or the first Pontrjagin class (however, only the Gauss-Bonnet
density will belong to the particular class of theories considered in
detail below). Let us point out that, in general theories depending on
derivatives of the curvature up to order $m$, the generalization of
eq.~(1) is:
\begin{displaymath}
  S_{\rm BH} =
   -2\pi\int_{\Sigma}\left(
   \frac{\partial L}{\partial R_{abcd}} -
   \nabla_{a_1}\frac{\partial L}
   {\partial \nabla_{a_1}R_{abcd}} + \cdots +(-1)^m
   \nabla_{(a_1\cdots}\nabla_{a_m)}\frac{\partial L}
   {\partial \nabla_{(a_1\cdots}\nabla_{a_m)}R_{abcd}}
   \right)
   \epsilon_{ab}\, \epsilon_{cd}\,
   \sqrt{h}\, d\Omega\, ,
\end{displaymath}
and thus all considerations above are qualitatively the same in this
case. Therefore we do not foresee any essential difficulty in
extending our results in a straightforward manner to more general
theories depending on derivatives of the curvature.}. Of course, the
expression~(\ref{kkdevaka}) {\em is} unambiguously defined when
evaluated on a spacelike cross-section of the event horizon of a
stationary black hole (cf. eq.~(\ref{WaldEntropy})), but not on
general spacelike surfaces, thus failing to fulfill requirement~a)
above. Below we briefly review what makes a BH horizon special in this
respect.

Our result for the $L_g(R)$-theories is nevertheless valid, since, in
four dimensions, one cannot build total derivatives out of $R$
alone. However, any definition of $\conhoya{C}$ in a general theory
will necessarily have to face this issue\footnote{In fact, we will
give below a general definition of $\conhoya{C}$ which is free of
these ambiguities, and we will see that it reduces to the
form~(\ref{kkdevaka}) for the particular case of theories with a
Lagrangian $L_g=L_g(R)$.}. Hence we see that, in a general theory, we
have to refine the ``naive'' definition of $\conhoya{C}$ so that it
has the following property:
\begin{itemize}
  \item[d)] Its definition has to be absent from any ambiguities. In
  particular, we require from $\conhoya{C}$ to be insensitive to the
  addition of a total derivative or a ``topological density'' to the
  Lagrangian.
\end{itemize}

Another nontrivial challenge when considering more general theories of
gravity is to find out which are the ``physical conditions'' to be
fulfilled in order to have property~c) above.
In particular, it is nontrivial to foresee what should be the
corresponding conditions of ``positivity of energy''. In a general
theory, one has to find how the generalization of
condition~(\ref{Condition}) comes about, if such condition is
physically meaningful, and if imposing such condition makes sense on
physical grounds. \\

This Section is devoted to the construction of a
$\conhoya{C}$-function that satisfies by construction
properties~a),~b) and~d) above. The proof of property~c) beyond the
case $L_g(R)$ is complicated in general. Nevertheless, we will see
that, in Lagrangians of the form $L_g=L_g(R,P,Q)$ (see
eq.~(\ref{LagrangianRPQ})), whenever the theory is known to be
physically meaningful\footnote{In particular we are talking about the
cancellation of ghosts in the spectrum --- see below.}, property~c)
above is also obeyed.

\subsection{Construction of the $\conhoya{C}$-function}
\label{sec:4.1}

Let us show now how one can define a function $\conhoya{C}$ fulfilling
all the requirements~a),~b) and~d) presented at the beginning of this
Section. In particular, let us start by discussing in detail points~a)
and ~d) above when the $\conhoya{C}$-function is taken to be of the
form~(\ref{kkdevaka}). That is, we start by analyzing the
generalization provided by formula~(\ref{WaldEntropy}) when evaluated
on a general spacelike surface~${\cal S}$ rather than just on a black
hole horizon, i.e.:
\begin{equation}
  \label{SBHgS}
  S_{\rm BH}[g,{\cal S}] = -2\pi\int_{\cal S}
  \frac{\p L}{\p R_{abcd}}
  \epsilon_{ab}\, \epsilon_{bc}\, \sqrt{h}\, d\Omega\, ,
\end{equation}
where, in the l.h.s., $g$ makes explicit reference to the background,
and ${\cal S}$ indicates the surface on which the functional $S_{\rm
BH}$ is to be evaluated. As discussed in the previous Section, this
expression is, in general, sensitive to the addition to the Lagrangian
of a total derivative depending upon the curvature. Consider however
the case of a stationary black hole spacetime. In such a case, it is
believed that the BH horizon is always a Killing horizon of some
Killing vector $\xi^a$ \fn{As it is well known, this fact was proved
in \cite{Hawking_2nd_law} for GR. While this fact has not been proved
for an arbitrary higher curvature theory of gravity, no
counterexamples are known to the best of our knowledge.  In
particular, this property automatically holds for static, spherically
symmetric black holes, as argued in~\cite{Wald:1993nt}.}.
In~\cite{Iyer:1994ys} it was shown that all additional terms arising
in $S_{\rm BH}$ from the addition of an exact form to the Lagrangian
are always proportional to $\xi^a$. Therefore, if the integration
surface is chosen to be the bifurcation surface, $\Sigma_{\rm bif}$,
of the Killing horizon (which is, by definition, the surface at which
$\xi^a$ vanishes), these contributions cancel\fn{Below we will
consider explicitly the addition to the Lagrangian of a topological
term (in particular, the addition of a Gauss-Bonnet density). We will
see that, in such a case, the final expression for the entropy {\em
does} get modified. However, this modification is just the shift of
the entropy by a constant.}.  Furthermore, the
authors of~\cite{Jacobson:1993vj} (see also~\cite{Iyer:1994ys}) showed
that:
\begin{equation}
   \label{ResultOfMJ}
   S_{\rm BH}[g, \Sigma_{\rm bif}] = S_{\rm BH}[g,\Sigma]\, ,
\end{equation}
where $\Sigma$ above is any arbitrary spacelike cross-section of the
horizon. As a result, $S_{\rm BH}[g, \Sigma]$ is always well defined,
but, on the other hand, it will be clearly ambiguous if the surface of
integration is taken to be an arbitrary surface ${\cal S}$. \\

The crucial point in this paper is to find a generalization of $S_{\rm
BH}[g,\Sigma]$ which is well defined on any arbitrary surface ${\cal
S}$. Moreover, we necessarily require such a generalization to reduce
to $S_{\rm BH}[g,\Sigma]$ when $g_{ab}$ describes a stationary BH
spacetime and the surface ${\cal S}$ is chosen to be a cross-section
$\Sigma$ of the BH horizon. This problem was essentially solved by
Iyer and Wald in~\cite{Iyer:1994ys}. In that reference they provided
an algorithm to define the entropy of a dynamical
(i.e. non-stationary) black hole. Such a case confronts the same
problems we have here since, in general, the event horizon will no
longer be a Killing horizon (a non-stationary black hole spacetime may
have no Killing vectors at all). Our solution to the problem at hand
is clearly inspired by Iyer and Wald's proposal for the entropy of a
dynamical black hole; however, it differs from it. For the interested
reader, we review their proposal in Appendix~\ref{Appendix} (the
remaining of the paper is however self-contained and does not require
the use of the results reviewed in the Appendix).

\subsubsection{General idea}
\label{GeneralIdea}

In order to provide a generalization of~(\ref{SBHgS}) which is well
defined on arbitrary spacelike surfaces, we proceed as follows. As
in~\cite{Iyer:1994ys}, instead of modifying the functional form of
$S_{\rm BH}[g,{\cal S}]$ given in (\ref{SBHgS}), we provide an algorithm
that deforms the spacetime metric $g_{ab}$ in the vicinity of ${\cal
S}$. We call $\conhoya{g}_{ab}$ the metric of this deformed
spacetime. This deformation is such that ${\cal S}$ becomes the
bifurcation surface of a bifurcate Killing horizon in the (artificial)
spacetime $\conhoya{g}_{ab}$. This will make the expression $S_{\rm
BH}[\conhoya{g},{\cal S}]$ to be automatically well defined (for
exactly the same reasons that make $S_{\rm BH}[g,\Sigma]$ to be well
defined at the horizon of a stationary black hole), hence having
fulfilled properties~a) and~d) above. On the other hand, the
particular deformation $g\rightarrow\conhoya{g}$ that we choose will
be such that properties~b) and~c) above will also be
obeyed\footnote{The main difference between our proposal and that
of~\cite{Iyer:1994ys} is that, if we deform the original spacetime in
the way proposed in~\cite{Iyer:1994ys}, then property~c) above does
not hold. See the Appendix for an explicit comparison.}. \\

The basic observation to find such a deformation $\conhoya{g}_{ab}$ is
the following. Any spacetime which, in certain coordinates
$(U,V,x^1,x^2)$, has a metric of the form:
\beqa
  \label{UV_metric}
  ds^2=F(UV,x^1,x^2)\, dU dV+g_{ij}(UV,x^1,x^2)\, dx^i dx^j\, ,
\eeqa
(where $F$ and $g_{ij}$ are arbitrary functions of $(x^1, x^2)$ and
the single combination $UV$) admits the following Killing vector:
\beqa \label{KillingUV} \xi=U\p_U-V\p_V\, ,  \eeqa
and has a bifurcate Killing horizon (the corresponding Killing vector
being $\xi$ above) at $UV=0$, with bifurcation surface (parametrized
by~$(x^1,x^2)$) at $U=V=0$. In the metric above, the null coordinates
$(U,V)$ (that we will define below in a precise manner) can be thought
as a generalization of the usual Kruskal coordinates of the
Schwarzschild black hole. In fact, the simplest spacetime with a
metric of the kind~(\ref{UV_metric}) is Schwarzschild spacetime. The
Killing vector~(\ref{KillingUV}) generates Lorentz boosts in the
subspace parametrized by~$(U,V)$. Therefore (following the
nomenclature used in~\cite{Iyer:1994ys}) we will refer to such metrics
as {\em boost-invariant} metrics\footnote{Note that defining $U=T+X$,
$V=T-X$, the metric~(\ref{UV_metric}) only depends on the
Lorentz-invariant combination $-T^2+X^2$.}. \\

Given any spacetime metric $g_{ab}$ and an arbitrary spacelike surface
${\cal S}$, the basic idea is the following. First, go to an
appropriate system of ``Kruskal coordinates'' $(U,V)$ (to be defined
below) in the space orthogonal to ${\cal S}$, parametrized in such a
way that ${\cal S}$ lies at $U=V=0$. In general (unless ${\cal S}$ is
already the bifurcation surface of a bifurcate Killing horizon), the
resulting metric will not be of the form~(\ref{UV_metric}). Therefore
the next thing to do will be to define a new (unphysical) spacetime
$\conhoya{g}_{ab}$ in a vicinity of $U=V=0$, such that
$\conhoya{g}_{ab}$ has the general form~(\ref{UV_metric}). Of course,
the deformation has to be such that $\conhoya{g}_{ab}$ is obtained
from $g_{ab}$ in a unique fashion (at least modulo ``gauge freedom''
--- see below). Let us next proceed to explain the construction of
$\conhoya{g}_{ab}$.

\subsubsection{Boost-invariant projection of $g_{ab}$ at ${\cal S}$}
\label{sec:BIProjection}

Consider an arbitrary spacetime $g_{ab}$ and an arbitrary spacelike
surface~${\cal S}$. As said, the first thing to do is to find an
appropriate system of null coordinates $(U,V)$ perpendicular to ${\cal
S}$, such that ${\cal S}$ is at $U=V=0$. For a completely generic
spacetime, a ``universal'' coordinate system, which is always possible
to define in a vicinity of ${\cal S}$, was introduced
in~\cite{Kay:1988mu,Iyer:1994ys} (we review this coordinate system in
the Appendix). It has the right property of automatically bringing the
metric to the form~(\ref{UV_metric}) whenever ${\cal S}$ is already
the bifurcation surface of a bifurcate Killing horizon.

However, the situation is somewhat simpler in the case of static
spacetimes. Since all examples considered below will be of this kind,
let us write here the explicit coordinate system that we will be using
in such cases. Consider a static spacetime parametrized as in
eq.~(\ref{SphSymmST}). In such a case, we define the following system
of ``Kruskal coordinates'' $(U,V)$ given by:
\begin{equation}
  \label{Txuskal}
  \begin{array}{rclrcl}
  u &=& \displaystyle t+\int {dr\over a(r)}\, , \hspace{2.5cm} &
  U &=& e^{u}\, , \\[.4cm]
  v &=& \displaystyle t-\int {dr\over a(r)}\, . &
  V &=& -e^{-v}\, .
  \end{array}
\end{equation}
In this coordinate system, the metric~(\ref{SphSymmST}) becomes:
\begin{equation}
  \label{StaticInTxuskal}
  ds^2 = \frac{a[r(UV)]}{UV}\, dUdV + b[r(UV)]\, d\Omega^2\, .
\end{equation}
Note that this metric has always the boost-invariant
form~(\ref{UV_metric}). Actually, if we have a regular event horizon
at $r=r_H$, we have that $a(r)\sim (r-r_H)$ near the horizon, and
therefore:
\begin{equation}
  r\rightarrow r_H\ \ \Leftrightarrow \ \ UV\rightarrow 0\, ,
\end{equation}
as desired\footnote{Since~(\ref{Txuskal}) always brings a static
metric to the form~(\ref{UV_metric}), it might seem that we will
always have a Killing horizon. Of course, this is not the case. If we
start with a metric without horizons, the region $UV=0$ will not be a
physical part of the spacetime. The example of $AdS$ considered below
will be an explicit example of this.}. Note that, in terms of the
original coordinates, the Killing vector~(\ref{KillingUV}) is
$\xi=\p_t$. \\

Remember that $\conhoya{C}$ is to be defined along outgoing null
geodesic flow. Therefore, in the particular case of static metrics (\ref{SphSymmST}) that we are considering,
the surfaces ${\cal S}$ we will be interested in will be spheres. An
arbitrary sphere ${\cal S}$ in the spacetime~(\ref{StaticInTxuskal})
will be located at $(U_0,V_0,x^i)$ with fixed $U_0$, $V_0$. Therefore
the next step is to shift coordinates:
\begin{equation}
  U \rightarrow (U+U_0)\, , \hspace{1cm}
  V \rightarrow (V+V_0)\, ,
\end{equation}
so that we get ${\cal S}$ at $U=V=0$. In general, since ${\cal S}$
will not be the bifurcation surface of a Killing horizon, the
resulting metric will not be boost-invariant there. Therefore we
define the {\em boost-invariant projection of $g_{ab}$ at ${\cal S}$},
denoted by $\conhoya{g}_{ab}$, as follows:
\begin{equation}
  \label{GeneralBI}
  \conhoya{g}_{ab} \equiv \sum_{n=0}^{N/2} (C_{ab})_n\, (UV)^n\, ,
\end{equation}
(where we have omitted the explicit dependence on the $x^i$), and
where $N$ is the order of the higher derivative of the metric
appearing in the Lagrangian. The coefficients $(C_{ab})_{n}$ above are
to be chosen as follows:
\begin{itemize}
  \item{First take $(C_{ab})_{0}$ to be equal to $g_{ab}$ at $U=V=0$,
  in order to make both metrics to coincide at ${\cal S}$.}
  \item{Next, choose the remaining coefficients $(C_{ab})_{n}$ such
  that all independent curvature invariants constructed out of
  $\conhoya{g}_{ab}$ that contribute to $S_{\rm BH}[\conhoya{g},{\cal
  S}]$ match those constructed out of $g_{ab}$ at $U=V=0$.}
\end{itemize}
The last point requires some explanation. Notice that, in the case of
a boost-invariant metric, the number of curvature invariants that will
enter in the final form of $S_{\rm BH}$ is always lower than in a
generic (not boost-invariant) case. This is because, for a metric
like~(\ref{GeneralBI}), the way in which ${\cal S}$ is embedded in
$\conhoya{g}_{ab}$ always makes both extrinsic curvatures of ${\cal
S}$ to vanish. This implies identities between the intrinsic curvature
invariants of ${\cal S}$ and those of $\conhoya{g}_{ab}$ at ${\cal
S}$, which will reduce the number of independent curvature invariants
appearing in the formula for the entropy.

In particular, one can derive the following relation (already used
in~\cite{Iyer:1994ys}, and which we will use below) relating the
intrinsic curvature of ${\cal S}$ with the curvature invariants of
$\conhoya{g}_{ab}$ at ${\cal S}$: let $k^a$ and $n^a$ be two null
vectors normal to ${\cal S}$ along the coordinate $U$ and $V$,
respectively, and normalized so that $k^an_a=-1$. Then, if ${}^{(2)}R$
is the intrinsic curvature of ${\cal S}$, the following identity
holds:
\begin{equation}
  \label{Curvatures}
  {}^{(2)}R = \Big(R - 2t^{ab}R_{ab} +
  t^{ac}t^{bd}R_{abcd}\Big)_{U=V=0}
\end{equation}
where
\beqa \label{t_ab} t_{ab}\equiv-k_an_b-n_ak_b \eeqa
is the induced metric in the space orthogonal to ${\cal S}$. Such
relations do not generically hold if the metric is not boost
invariant\fn{The precise statement is that the relation
(\ref{Curvatures}) holds at a given point if, at that point, the
extrinsic curvatures with respect to $k^a$ and $n^a$ ($K_{ab}\equiv
h_a^{\ i}h_b^{\ j}\nabla_ik_j$ and $N_{ab}\equiv h_a^{\ i}h_b^{\
j}\nabla_in_j$, with $h_{ab}=g_{ab}+k_an_b+n_ak_b$ the induced metric
in ${\cal S}$) satisfy $K_a^{\ b}N_b^{\ a}=K_c^{\ c}N_d^{\ d}$.}. \\

Let us finish this Section with the following comments:
\begin{itemize}
  \item{Note that the boost-invariant projection of $g_{ab}$ at ${\cal
  S}$ does not always exist. This is because we have at our disposal a
  maximum number of coefficients to fit and, in principle, it is not
  ensured that they can be chosen to match all possible invariants
  contributing to $S_{\rm BH}[\conhoya{g},{\cal S}]$\footnote{Of course, the
  value of $N$ in the expansion~(\ref{GeneralBI}) can be higher than
  the order of the highest derivative of the metric appearing in the
  Lagrangian. However, in such a case, all ``extra'' coefficients are
  redundant since they do not contribute to the curvature at $U=V=0$
  (this is what we meant by ``gauge freedom'' in
  Section~\ref{GeneralIdea}).}. However, we will see that, in the case
  of the theories $L_g(R,P,Q)$, the number of curvature invariants
  entering in the final expression for $S_{\rm
  BH}[\conhoya{g},{\cal S}]$ precisely coincides with the number of
  coefficients at our disposal {\em if} the theory is ghost-free, thus
  making the construction of $\conhoya{g}_{ab}$ always possible. We
  therefore expect that, in a general theory of gravity, cancellation
  of the non-physical degrees of freedom will make the election of a
  boost-invariant projection of the original metric always possible.}
  \item{Note that {\em any} metric of the form~(\ref{GeneralBI}) is
  boost-invariant (with ${\cal S}$ being the bifurcation surface of a
  Killing horizon) for whatever choice of the coefficients
  $(C_{ab})_n$. Iyer and Wald's proposal in~\cite{Iyer:1994ys} is in
  fact a similar way of deforming the original spacetime, but with a
  different prescription to choose these coefficients (see
  Appendix). Our particular choice will be justified a posteriori,
  since we will see that it is precisely this choice what makes
  property~c) at the beginning of this Section to be satisfied.}
  \item{From the technical point of view, in this Section we have
  focused in the case of static spacetimes. We have done this just
  for simplicity of the exposition, but our results remain valid (and
  our prescription for defining $\conhoya{g}_{ab}$ unchanged)} for
  general spacetimes. The only technical difference is that, in a
  general case, the Kruskal coordinates defined in~(\ref{Txuskal})
  will not be appropriate. However, one can always define the $(U,V)$
  coordinates as explained in~\cite{Kay:1988mu,Iyer:1994ys} (see
  Appendix), all the rest being exactly the same.
\end{itemize}

\subsubsection{General definition of $\conhoya{C}$}

Finally, let us provide here our definition of the
$\conhoya{C}$-function in a general theory of gravity. Given a
solution $g_{ab}$ of the equations of motion, and given a spacelike
surface ${\cal S}$, we define $\conhoya{C}$ as:
\begin{equation}
  \label{ConhoBernarda}
  \conhoya{C}[g,{\cal S}] \equiv
  S_{\rm BH}[\conhoya{g}, {\cal S}] =
   -2\pi\int_{\cal S}
   \frac{\p L(\conhoya{g})}{\p \conhoya{R}_{abcd}}\,
   \conhoya{\epsilon}_{ab}\,
   \conhoya{\epsilon}_{cd}\, \sqrt{h}\, d\Omega\, ,
\end{equation}
where $\conhoya{g}_{ab}$ is the boost-invariant projection of $g_{ab}$
at ${\cal S}$, as defined in the previous Section. Notice that such a
$\conhoya{C}$-function automatically satisfies properties~a), b)
and~d) enumerated at the beginning of this Section. Properties~a)
and~d) are satisfied by construction. The fact that $\conhoya{C}$
equals the entropy of a stationary black hole when evaluated on a
spacelike surface $\Sigma$ of its event horizon is maybe less
evident. However, this fact follows from the definition of
$\conhoya{g}_{ab}$: remember that this metric is chosen so that every
invariant contributing to the final form of $S_{\rm
BH}[\conhoya{g},{\cal S}]$ equals those of the original spacetime at
${\cal S}$. At the bifurcation surface $\Sigma_{\rm bif}$ of a regular
event horizon, the spacetime metric $g_{ab}$ will be automatically
boost-invariant (cf.~\cite{Kay:1988mu} or last Section for the
particular case of static spacetimes). Therefore, the same relations
like~(\ref{Curvatures}) will hold both for $g_{ab}$ and
$\conhoya{g}_{ab}$ at $\Sigma_{\rm bif}$. This will make appear the
same independent curvature invariants in $S_{\rm
BH}[\conhoya{g},\Sigma_{\rm bif}]$ as in $S_{\rm BH}[g,\Sigma_{\rm
bif}]$. Since $\conhoya{g}_{ab}$ is chosen such that those are equal,
then one necessarily has:
\begin{equation}
  \conhoya{C}[g,\Sigma_{\rm bif}] = S_{\rm BH}[g,\Sigma_{\rm bif}]\, .
\end{equation}
This, together with the result~(\ref{ResultOfMJ})
of~\cite{Jacobson:1993vj}, implies that:
\begin{equation}
  \conhoya{C}[g,\Sigma] = S_{\rm BH}[g,\Sigma]\, .
\end{equation}
Below we will see an explicit example of this in the case of the
cosmological horizon of de Sitter space. \\

Summarizing, we have been able to construct an algorithm that allows
to define a $\conhoya{C}$-function in theories of higher curvature
gravity, subject to properties a), b) and d) above. However, we have
not shown that property~c) (namely, that $\conhoya{C}$ is an
non-decreasing function along outgoing radial null geodesics)
holds. As already emphasized, this is a nontrivial task in a
completely generic theory and in an arbitrary background; not only
because of the technical difficulty involved in the calculations, but
also due to the fact that the physical consistency requirements
necessary to prove condition~c) are not known in general. Therefore,
at this point, we can only leave our construction as a proposal whose
validity must in principle be tested in a case by case scenario. In
any case, we provide in the following Section the explicit
construction of the $\conhoya{C}$-function in a particular case:
theories whose Lagrangian is given by eq.~(\ref{LagrangianRPQ}) in
maximally symmetric backgrounds, where the physical consistency
conditions have been recently worked out~\cite{Navarro:2005da}. We
will see that in these cases property~c) above is indeed satisfied.

\section{$\conhoya{C}$ in theories of the form $L_g=L_g(R,P,Q)$}

We will devote this Section to illustrate the construction of the
$\conhoya{C}$-function in a particular case: maximally symmetric
backgrounds in theories where the Lagrangian is of the form given by
eq.~(\ref{LagrangianRPQ}), i.e. $L_g=L_g(R,P,Q)$ where $P\equiv
R_{ab}R^{ab}$ and $Q\equiv R_{abcd}R^{abcd}$. We will first review the
consistency conditions for these backgrounds in these kind of
theories, and then we will compute $\conhoya{C}$ following the
prescription given in Section~\ref{sec:ConstructionOfTheConho}. We
will show how this function satisfies property~c) above whenever the
physical consistency conditions are met.

\subsection{Consistency of the $L_g(P,Q,R)$ theories}
\label{PropertiesOfLPQR}

In general, it is known that higher derivative couplings introduce new
degrees of freedom in the theory. The case of theories linear in $R$,
$R^2$, $P$ and $Q$ in four dimensions was studied a long time ago by
Stelle~\cite{Stelle:1976gc}. It is found that the spectrum of these
theories includes, in addition to the massless graviton, a massive
scalar (which can be tachyonic or not) and a massive spin-2 field
which is always a ghost. \\

Gravity theories with a gravitational Lagrangian of the form
$L_g(R,P,Q)$ (see eq.~(\ref{LagrangianRPQ})) have been studied
recently in~\cite{Navarro:2005da}. Their consistency and stability of
a given background depends of course of the particular function $L_g$
chosen and on the particular background considered.  Unfortunately,
the perturbative expansion of these theories about a generic
background is not known. However, on maximally symmetric backgrounds,
it can be shown that the spectrum and the perturbative expansion of
the $L_g(R,P, Q)$-theory are equivalent those of the following
Lagrangian:
\begin{equation}
  \label{LRC}
  L_g(R,C) = -2\Lambda + \delta R + \frac{1}{6 m_0^2} R^2
           + \frac{1}{2 m_2^2} C_{abcd} C^{abcd}\, ,
\end{equation}
about the same maximally symmetric background. $C_{abcd}$ above
denotes the Weyl tensor, and the parameters $\delta$, $m_0^2$ and
$m_2^2$ appearing in $L_g(R,C)$ are calculated from the original
Lagrangian $L_g(R,P,Q)$ as follows~\cite{Navarro:2005da}:
\begin{equation}
  \label{CoefficientsOfLRC}
  \begin{array}{rcl}
  \delta & = & \Big(
  L_R - R L_{RR} - R^2 ( L_{RP} + \frac{2}{3} L_{RQ} )
  - R^3 ( \frac{1}{4} L_{PP} + \frac{1}{9} L_{QQ}
        + \frac{1}{3} L_{PQ})
  \Big)_0 \, , \\[.4cm]
  m_0^{-2} & = & 3\, \Big(
  L_{RR} + \frac{2}{3}(L_P + L_Q) +
  R ( L_{RP} + \frac{2}{3} L_{RQ} )\ + \\[.4cm]
  && \ \
  +\ R^2 ( \frac{1}{4} L_{PP} +
           \frac{1}{9} L_{QQ} + \frac{1}{3} L_{PQ} )
  \Big)_0\, , \\[.4cm]
  m_2^{-2} & = & \Big( L_P + 4 L_Q \Big)_0\, .
  \end{array}
\end{equation}
The subscript ``$0$'' in the r.h.s. denotes evaluation in the
maximally symmetric background, and $L_R$, $L_{RP}$, etc. denote the
corresponding partial derivatives of $L_g$ with respect to $R$, $P$
and $Q$. The value of the cosmological constant $\Lambda =
-\frac{1}{2} (L_g + \cdots)_0$ will not be needed in what follows. The
value of $\Lambda$ is related to the curvature of the maximally
symmetric space by the equation of motion:
\beqa
  (2L_Q+3L_P)R^2+6L_RR-12 L_g=0\, .
  \label{eomLRPQ}
\eeqa

The theory~(\ref{LRC}) falls into the class of theories studied
in~\cite{Stelle:1976gc}\footnote{This is due to the fact that any
theory with $L_g$ linear in $R$, $R^2$, $P$ and $Q$ can be put into
the form~(\ref{LRC}) by using the identity:
\begin{displaymath}
  C_{abcd}C^{abcd} = Q - 2P + \frac{1}{3}R^2\, ,
\end{displaymath}
and by using the fact that, in four dimensions, the Gauss-Bonnet
density
\begin{displaymath}
  GB=R^2-4P+Q=C_{abcd}C^{abcd}-2P+\frac{2}{3}R^2
\end{displaymath}
does not contribute to the equations of motion.}, and therefore the
spectrum is just as explained above. Actually, the masses of the
scalar and the spin-2 field are proportional, respectively, to $m_0$
and $m_2$. Thus, while it is true that theories of the kind
$L_g(R,P,Q)$ have, in general, a ghost-like spin-2 field in the
spectrum, we see that we can decouple the ghost if we have
$(L_P+4L_Q)_0=0$. In particular, all theories with a Lagrangian of
the form:
\begin{equation}
  \label{RestrictedLPQR}
  L_g(R,P,Q)=L_g(R,T=Q-4P)\,
\end{equation}
will be automatically ghost-free.  In~\cite{Navarro:2005da} it was
shown that this condition to have a ghost-free theory remains valid in
the case of a FRW cosmological background. \\

So the first thing we have to do is to make the theory ghost-free.  In
order to achieve this, we restrict ourselves to Lagrangians of the
form~(\ref{RestrictedLPQR}). Then the ghost decouples from the theory
and, further, $L_g(R,C)$ in~(\ref{LRC}) reduces to:
\begin{equation}
  L_g(R,C) = -2\Lambda + \delta R + \frac{1}{6m_0^2} R^2\, ,
\end{equation}
which is just of the form $L_g=L_g(R)$ studied in
Section~\ref{LRTHeories} \footnote{The cosmological constant can be
thought as a contribution to the matter Lagrangian satisfying the null
energy condition}. There we showed that a consistency requirement in
this kind of theories is (see also~\cite{Jacobson:1995uq}):
\begin{equation}
  \frac{\partial L_g(R,C)}{\partial R} > 0\, .
\end{equation}
Using~(\ref{CoefficientsOfLRC}) and the no-ghost condition we see
that this translates into:
\begin{equation}
  \label{ConsistencyReq}
  L_R - 2 L_T R > 0\, ,
\end{equation}
where the functions appearing above are functions only of $R$ and
$T$, and these invariants are those of the maximally symmetric
background considered.

\subsection{$\conhoya{C}$-function in maximally symmetric backgrounds}

Let us see what we get in these theories from our general
considerations to find the $\conhoya{C}$-function. First we observe
that, for any metric $g_{ab}$, the functional form of $S_{\rm
BH}[g,{\cal S}]$ in~(\ref{SBHgS}) for a generic Lagrangian
$L_g(R,P,Q)$ is given by:
\begin{equation}
  \label{SBHInLRPQ}
  S_{\rm BH}[g,{\cal S}] = 4\pi \int_{\cal S}
  (L_R + L_P\, R^{ab}t_{ab}
   + 2 L_Q\, R^{abcd}t_{ac}t_{bd}\, )\, \sqrt{h}\, d\Omega\, ,
\end{equation}
where we recall that $t_{ab}$ is the induced metric on the spacetime
orthogonal to ${\cal S}$ (see eq.~(\ref{t_ab})). Now we wish to
evaluate $\conhoya{C}[g,{\cal S}]$, eq.~(\ref{ConhoBernarda}). The
first thing to do is to find $\conhoya{g}_{ab}$, the boost-invariant
projection of $g_{ab}$. For these theories, $\conhoya{g}_{ab}$ will
always be a metric of the form~(\ref{GeneralBI}) with $N=2$, since the
Lagrangian does not include any derivative of the metric of degree
greater than two. Maximally symmetric spaces can be always
parametrized as in~(\ref{SphSymmST}). Therefore, after we switch to
Kruskal coordinates~(\ref{Txuskal}), we will always get a boost
invariant Ansatz of the kind:
\begin{equation}
  \label{conhooo}
  d\conho{s}^2 =
  \Big( (C_{UV})_0 + (C_{UV})_1\, UV \Big)\, dU dV +
  \Big( (C_{\Omega})_0 + (C_{\Omega})_1\, UV \Big)\, d\Omega^2\, ,
\end{equation}
where the coefficients $(C_{ab})_n$ will be determined below. Now, as
stated in Section~\ref{sec:ConstructionOfTheConho}, any metric of the
form~(\ref{GeneralBI}), as the metric above, will always satisfy the
identity~(\ref{Curvatures}). Additio\-nally, if we restrict ourselves
to the ghost-free theories~(\ref{RestrictedLPQR}), then we will have
$L_P=-4L_Q$. Putting these two conditions together, we can see
from~(\ref{SBHInLRPQ}) that:
\begin{equation}
  \label{SBHBI}
  S_{\rm BH}[\conhoya{g},{\cal S}] = 4\pi\, \int_{\cal S}
  \Big((\conhoya{L_R}) - 2(\conhoya{L_T})\conhoya{R}
  \Big)\sqrt{h}\, d\Omega
  + 8\pi\, \int_{\cal S} (\conhoya{L_T})\, {}^{(2)}R\,
  \sqrt{h}\, d\Omega\, ,
\end{equation}
where tilded symbols denote the corresponding quantities evaluated in
$\conhoya{g}_{ab}$. Now, notice that the final expression for whatever
choice of $\conhoya{g}$ depends on just two curvature invariants:
$\conhoya{R}$ and $\conhoya{T}$. Following
Section~\ref{sec:ConstructionOfTheConho}, we now choose the values of
the coefficients in~(\ref{conhooo}) so that:
\begin{itemize}
  \item{The $n=0$ coefficients have to be chosen so that
  $\conhoya{g}_{ab}=g_{ab}$ at ${\cal S}$.}
  \item{The $n=1$ coefficients have to be chosen so that
  $\conhoya{R}=R$ and $\conhoya{T}=T$ at ${\cal S}$.}
\end{itemize}
As already emphasized, terms of order $\sim (UV)^2$ or higher will not
change the curvature at $U=V=0$. This means that we have the same
number of coefficients to adjust in~(\ref{conhooo}) than the number of
equations derived from the conditions above. Hence we see that, in
general, a boost-invariant projection such that $\conhoya{R}=R$ and
$\conhoya{T}=T$ will always exist. Therefore we get a
$\conhoya{C}$-function given by:
\begin{equation}
  \label{GeneralCInLRPQ}
  \conhoya{C}[g,{\cal S}] = 4\pi\, \int_{\cal S}
  (L_R - 2L_T\, R)\sqrt{h}\, d\Omega
  + 8\pi\, \int_{\cal S} L_T\, {}^{(2)}R\,
  \sqrt{h}\, d\Omega\, .
\end{equation}
Integrating the above expression on a sphere of radius $r_0$ (recall
that for metrics of the form (\ref{SphSymmST}) the surface ${\cal S}$
is a sphere), we get:
\begin{equation}
  \label{FinalCInMSB}
  \conhoya{C} = 16\pi^2 b^2\, (L_R - 2L_T R)\, r_0^2 + 64\pi^2\, L_T\, ,
\end{equation}
where the expressions in the r.h.s. only depend on $R$ and $T$, the
curvature scalars of the maximally symmetric spacetime
$g_{ab}$. Notice that, in general, eq.~(\ref{GeneralCInLRPQ}) reduces
to the $\conhoya{C}$-function of Section~\ref{LRTHeories} for theories
whose Lagrangian depends only on the Ricci scalar.

\subsubsection{Properties of $\conhoya{C}$}

By construction, the expression~(\ref{FinalCInMSB}) has to be well
defined. In particular, it has to be insensitive to the addition of a
topological term to the Lagrangian\fn{As mentioned already in
footnote~\ref{Footnote4}, in four dimensions we have two possible
topological terms built up of local integrals of the curvature: the
first Euler class (i.e. the Gauss-Bonnet density) and the first
Pontrjagin class. The latter, however, does not belong to the
$L_g(R,P,Q)$ theories considered here.}. It is straightforward to
check that the addition of a Gauss-Bonnet term ($GB=R^2-4P+Q$) to the
Lagrangian just shifts $\conhoya{C}$ by a constant (proportional to
the Euler number of ${\cal S}$)\footnote{Hence we see that
$\conhoya{C}$ is not exactly invariant against the addition of a
Gauss-Bonnet density; however (as in the entropy of a physical system)
such a constant is clearly immaterial.}.\\

Second, if we were in a black hole spacetime, $\conhoya{C}$ should
equal the entropy of the black hole when evaluated at the
horizon. However, the result above is for maximally symmetric
spacetimes. Below we will consider explicitly the case of de~Sitter
space in the static patch. In this case, we will verify that, when
$\conhoya{C}$ is evaluated at the cosmological horizon it exactly
equals its entropy as when computed from
eq.~(\ref{WaldEntropy}). Moreover, notice also that $\conhoya{C}$
reduces to the entropy formula for Lovelock gravity obtained
in~\cite{Iyer:1994ys,Jacobson:1993xs}.\\

Finally, note that all the curvature scalars are constant in a
maximally symmetric space, and hence this function is a non-decreasing
function of $r_0$ if and only if the theory is ghost-free (see
eq.~(\ref{ConsistencyReq})). Note that this is a natural
generalization of the ``positivity of energy''
condition~(\ref{Condition}) for the $L_g(R)$-theories (which, in turn,
was also a reasonable generalization of the null energy condition ---
see the comment below eq.~(\ref{EffectiveGN}) and
footnote~\ref{FNOnEffectiveGN}). In fact, condition~(\ref{Condition})
can be interpreted as a no-ghost condition for the graviton, since it
means that the graviton kinetic term has to carry the right sign. Here
such kind of condition simply extends to the extra degrees of freedom
appearing in the theory.

We could have expected to obtain another condition: the surviving
scalar in the spectrum being non-tachyonic. The fact that this
condition is not needed to have a well-behaved $\conhoya{C}$-function
should not come as a surprise. A tachyon in the spectrum means an
unstable background, examples of which exist already in GR;
nevertheless in GR a $\conhoya{C}$-function exists and the Second Law
holds irrespective of the stability of the background. \\

To end our discussion, let us next consider explicitly the cases of
de~Sitter and Anti~de~Sitter space.


\subsubsection{de Sitter Space}

This case is interesting since we have a cosmological horizon. From a
technical point of view, a cosmological horizon is the same as an
event horizon, so we will be able to check explicitly that our
$\conhoya{C}$-function equals its real entropy when evaluated at the
cosmological horizon.

The metric of $dS$-space in the static patch can be written as:
\begin{equation}
  ds^2 = b^2\Big( -(1-r^2)\, dt^2 + (1-r^2)^{-1}\, dr^2 +
  r^2\, d\Omega^2\Big)\, ,
\end{equation}
with $0 \leq r \leq 1$, $-\infty \leq t \leq \infty$; the cosmological
horizon is at $r=1$. First we have to go to Kruskal coordinates. These
are given by:
\beqa
  U=e^t\sqrt{{1-r\over 1+r}},
  \hspace{2cm} V=-e^{-t}\sqrt{{1-r\over 1+r}}.
\eeqa
These coordinates only cover the patch $\{U\geq 0\} \cup \{V\leq 0\}$
of $dS$ space. We can trivially extend the coordinates to the whole
range $-\infty \leq U,V \leq \infty$ in such a way that $r,t$ are
defined in each one of the patches as:
\beqa r={1+UV \over 1-UV},
  \hspace{2cm} t=\oh\log{\left|{U\over V}\right|}.
\eeqa
The metric is given, for all $U,V$, by:
\begin{equation}
  \label{KruskaldS}
  ds^2 = b^2\left( -\frac{4}{(1-UV)^2}\, dU dV
         + \left(\frac{1+UV}{1-UV}\right)^2\, d\Omega^2\right)\, .
\end{equation}
These coordinates cover all of de Sitter space. The horizon lies at
$UV=0$. Notice also that this metric is a function of the single
combination $UV$; therefore, at the bifurcate horizon $U=V=0$ (and
only there), the metric is automatically boost-invariant. In fact, the
vector:
\beqa
  \xi = U\p_U - V\p_V
  \label{xi2}
\eeqa
is a Killing vector of this metric, and vanishes at $U=V=0$. \\

Next we shift the coordinates in order to have an arbitrary sphere of
radius $r_0$ at $U=V=0$. We make the shift $U \rightarrow (U+U_0)$, $V
\rightarrow (V+V_0)$, so that $U=V=0$ becomes a sphere of
radius\fn{Note that if the point $(U_0,V_0)$ is located in the
initially considered patch, so that $U_0>0$, $V_0<0$, then $r_0$ is
always smaller than 1.}:
\begin{equation}
  r_0^2 = \left(\frac{1+U_0V_0}{1-U_0V_0}\right)^2\, .
\end{equation}
The metric in terms of the shifted coordinates is given by
\begin{equation}
  \label{KruskaldSshifted}
  ds^2 = b^2\left( -\frac{4}{\left(1-(U+U_0)(V+V_0)\right)^2}\, dU dV
         + \left(\frac{1+(U+U_0)(V+V_0)}{1-(U+U_0)(V+V_0)}\right)^2\,
         d\Omega^2\right)\, .
\end{equation}
Notice that this metric is not of the boost-invariant form
eq.~(\ref{UV_metric}). In fact, $\xi$ in (\ref{xi2}) is no longer a
Killing vector. However, it will be a Killing vector of its boost
invariant projection at the surface $U=V=0$ (i.e.  $r=r_0$ in some
patch), and moreover this surface will become the bifurcation surface
of a Killing horizon in the $\conhoya{g}_{ab}$-space. \\

Let us check explicitly that such a boost invariant projection exists
for de Sitter space.  The Ansatz for $\conhoya{g}_{ab}$ is given in
eq.~(\ref{conhooo}). Next have to impose the conditions stated in
Section~\ref{sec:ConstructionOfTheConho} to get the correct values for
the coefficients $(C_{ab})_n$. In order to match with the actual $dS$
metric~(\ref{KruskaldSshifted}) at $(U,V)=(0,0)$ we must impose
\beqa
  (C_{UV})_0=-(1+r_0)^2, \hspace{2cm} (C_{\Omega})_0=r_0^2.
\eeqa
The other two coefficients are adjusted so that
$\conhoya{R}=R_{dS}=12/b^2$ and $\conhoya{T}=T_{dS}=-120/b^4$. We get
two possible solutions for each coefficient given by:
\beqa
  (C_{UV})_1&=&{1\over r_0^2}(1+r_0)^4\left({5\over 2}-3r_0^2\pm
  \sqrt{6}|1-r_0^2|\right)\\ (C_\Omega)_1&=&(1+r_0)^2\left(1\pm
  {\sqrt{6}\over2}|1-r_0^2|\right)
\eeqa

Finally, using~(\ref{GeneralCInLRPQ}) and integrating on the sphere of
constant radius $r_0$, we get a $\conhoya{C}$-function given by
eq.~(\ref{FinalCInMSB}), where the r.h.s. is to be evaluated in
de~Sitter space. It can be explicitly checked that $\conhoya{g}_{ab}$
at $r_0=1$ exactly matches the series expansion of the de~Sitter
metric~(\ref{KruskaldS}) up to second order.  Therefore, if
$\conhoya{C}$ is evaluated at the cosmological horizon $r_0=1$, it
will equal its entropy as computed from eq.~(\ref{WaldEntropy}).

\subsubsection{Anti de Sitter Space}
\label{sec:AdS}

The metric of $AdS$ can be written as:
\begin{equation}
  ds^2 = b^2\Big( -(1+r^2)\, dt^2 + (1+r^2)^{-1}\, dr^2 +
  r^2\, d\Omega^2\Big)\, ,
\end{equation}
with $0\leq r \leq \infty$ and $-\infty \leq t \leq \infty$. The
relevant curvature scalars curvature are given by $R=-12/b^2$ and
$T=120/b^4$. These coordinates already cover the whole of $AdS$ space.
Next we define Txuskal coordinates by:
\begin{equation}
  \begin{array}{rclrcl}
  u &=& \displaystyle t+\arctan r\, , \hspace{2.5cm} &
  U &=& e^{u}\, , \\[.4cm]
  v &=& \displaystyle t-\arctan r\, . &
  V &=& -e^{-v}\, .
  \end{array}
\end{equation}
The metric becomes:
\beqa
  ds^2=b^2\left[{1\over UV}\sec^2\left(\oh \log(-UV)\right)dUdV
  +\tan^2\left(\oh\log(-UV)\right)d\Omega^2\right].
\eeqa
This metric is of the boost invariant form eq.~(\ref{UV_metric}), so
it apparently has a bifurcate horizon at $(U,V)=(0,0)$, that does not
exist in $AdS$ space. In fact, this ``horizon'' is just an artifact of
the change of coordinates, and the region $UV=0$ is unphysical and it
is not covered by the metric above. Notice that the relation between
the original coordinate $r$ and $(U,V)$ is:
\beqa
  UV=-\exp(2\arctan r) \label{uuve}
\eeqa
and hence $UV \in (-e^\pi,-1]$. It could be argued that other branches
of the arctan should be used. Taking a different branch of the
$\arctan$ changes the relation (\ref{uuve}) by
\beqa
  UV=-\exp(2\arctan r)\exp(2\pi n)
\eeqa
for some integer $n$. Now, we can see that this integer must be minus
infinity to reach $UV=0$, therefore we see that the ``horizon''
generated at $UV=0$ is definitely an artifact of the change of
coordinates and is outside $AdS$ space. \\

Next we proceed analogously to the $dS$ case. We shift the coordinates
as $(U,V)\to(U+U_0,V+V_0)$ (taking into account that
$U_0V_0\in(-e^{\pi},-1]$, in order to remain in $AdS$ space\fn{Or
$e^{\pi n}U_0V_0\in(-e^{\pi},-1]$ for some $n \in {\bf Z}$ for a
different branch of the arctan, considering also branches with negative $r$.}). The surface $U=V=0$ becomes a
sphere of radius:
\begin{equation}
  r_0^2 = \tan^2\left(\frac{1}{2}\log(-U_0V_0)\right)\, .
\end{equation}
Next we compute the boost-invariant projection $\conhoya{g}_{ab}$ at
$U=V=0$. Again, we impose that this metric coincides with that of
$AdS$ at $U=V=0$, and that $\conhoya{R}=R_{AdS}$ and
$\conhoya{T}=T_{AdS}$ at $U=V=0$. It can be explicitly checked that
this metric is given by~(\ref{conhooo}) with:
\beqa
(C_{UV})_0&=&-e^{-2\arctan r_0}(1+r_0^2),\nonumber\\
(C_{UV})_1&=&{e^{-4 \arctan r_0} (1+r_0^2)^2\over r_0^2} \left({5\over
2}+ 3r_0^2 \pm \sqrt{6} \sqrt{1+2r_0^2+11r_0^4}\right) \nonumber\\
(C_\Omega)_0&=&r_0^2,\\ (C_\Omega)_1&=&e^{-2\arctan r_0}(1+r_0^2)
\left(1\pm \sqrt{6} \sqrt{1+2r_0^2+11r_0^4}\right)\nonumber
\eeqa
Finally, the $\conhoya{C}$-function is given by
eq.~(\ref{FinalCInMSB}), where the r.h.s. is now to be evaluated in
$AdS$. \\

As already mentioned in the Introduction, let us stress here that we
are by no means claiming that the above function is the holographic
dual of a true field theory c-function (since, for $AdS$, such a
function should to be a constant). In order to address Questions~{\bf
1} and~{\bf 2} in the Introduction, the only property that we had to
demand from our ``c-function'' $\conhoya{C}$ was that it has to be
non-decreasing along outgoing null geodesic flow. This is exactly what
we got.

However, let us mention that, even without using any considerations on
the dual CFT, one would maybe expect that any geometric function in
$AdS$ should be independent of the radial coordinate, due to the
well-known ``scale invariance of Anti de Sitter space''. Let us recall
here that the latter is a property of the Poincar\'e patch of $AdS$,
but not a property of global $AdS$ (whose metric is the one that we
have used here\footnote{We wish to thank Ofer Aharony for pointing
this out to us.}). It is easy to see that had we constructed
the $\conhoya{C}$-function from the metric in the Poincar\'e patch
\beqa
ds^2=b^2\left(z^2(-d\tau^2+dx^2+dy^2)+{dz^2\over z^2}\right)
\eeqa
with $-\infty<\tau,x,y<\infty$, $0<z<\infty$,
we would have found $\conhoya{C}\sim z^2 ~dx\wedge dy$, that is
invariant under the symmetry $(\tau,x,y,z)\to (a \tau,ax,ay,a^{-1}z)$ (regardless of the fact of $\conhoya{C}$
being the holographic dual of a field theory c-function or not). Note
however that these are not good coordinates to address issues like the entropy contained in a given region of spacetime,
since there are no closed, compact spacelike 2-surfaces for
constant $\tau$ and $z$ in these coordinates.

\section{Conclusions}

In this paper we have explored to which extent the possibility of
establishing a well-defined holographic bound, as well as the Second
Law of black hole mechanics, extend to general theories of gravity
with higher curvature interactions. As we have shown, these two issues
seem to imply each other via the existence of a ``c-function'', i.e. a
never decreasing function along outgoing null geodesic flow. \\

It would be nice to explore the implications that the general
holographic bound that we propose here may have along the lines of
the covariant entropy bound of~\cite{Bousso:1999cb}. In order to
have well defined holographic screens, it seems crucial to prove
that a suitable generalization of the focussing theorem holds for 
$\conhoya{\vartheta}\equiv d\log{\conhoya{C}}/d\lambda$. We did not
attempt to do this in this paper, but it seems that such a
generalization should hold. After all, the focussing theorem plays a
crucial role in GR in the proof of the Second Law and in the proof
of all singularity theorems. In a sense, a convenient generalization
of Raychaudhuri equation and the focussing theorem could be taken as
a good starting point to define what a good theory of gravity should
be (including its recognition as
a holographic theory). \\

From a more technical point of view, it would be interesting to
understand the significance of the boost invariant projection
$\conhoya{g}_{ab}$ defined in Section~\ref{sec:BIProjection}. This
projection is central in our construction in order to arrive to the
final expression for the $\conhoya{C}$-function. However, we do not
understand the real significance (if any) of this spacetime. After
all, such a spacetime looks in the end like quite a spurious object
(having, in addition, a lot of ``gauge freedom'') which might well be
not much more than a mere artifact in order to obtain expressions
like~(\ref{Curvatures}), which seem to play a crucial form in the
final form of~$\conhoya{C}$. Also, the fact that we have to go to a
special coordinate system to find $\conhoya{g}_{ab}$ is somewhat
unsatisfactory. It would be very nice to understand the covariant
meaning of the boost invariant projection that we define. \\

Finally, let us mention here that an interesting corollary of our
definition of $\conhoya{C}$ is that it constitutes, by itself, a
possible candidate to define the entropy of a dynamical black hole,
since it satisfies by construction all the properties required
in~\cite{Iyer:1994ys}. However, it will in general differ from the
proposal of Iyer and Wald (see the Appendix). It would be nice to
check the implications of this. \\

Our results are far from general. However, we believe that the ideas
presented and developed in this paper should be of quite general
applicability. It would be very interesting to understand them in
depth and to check our proposal in more general theories and more
general backgrounds.

\section*{Acknowledgements}

We would like to thank Barak Kol, Fernando Quevedo and Aninda Sinha for comments on the manuscript, Sean Hartnoll and Alberto Ramos for useful conversations,
Ignacio Navarro for conversations and for helping us much in
understanding the consistency of gravity theories with higher
curvature couplings, and finally Cdo. Jorge Conde for useful comments on
the manuscript and for very useful suggestions on the notation of the paper.
We also want to thank the ``Bar-Cafeter\'{\i}a Mozar''
in Madrid for hospitality during the early stages of this project. D.C. also thanks the Weizmann Institute of Science for hospitality during the completion of
this work.

The work of D.C. is supported by the University of Cambridge. The work of E.~L.-T. has been supported by a Marie Curie Fellowship
under contract MEIF-CT-2003-502349, the Spanish Grant BFM2003-01090,
The Israel-US Binational Science Foundation, the ISF Centers of
Excellence Program and Minerva.

\appendix

\section{Iyer and Wald proposal for the entropy of a dynamical black hole}
\label{Appendix}

As mentioned in the body of the paper, our definition of the
$\conhoya{C}$-function is based on the proposal of Iyer and
Wald~\cite{Iyer:1994ys} to define the entropy of a non-stationary
black hole. Such a problem faces the same kind of difficulties that we
had to face in order to extend the entropy formula
eq.~(\ref{WaldEntropy}) to general surfaces which need not be a
cross-section of a Killing horizon (since the event horizon of a
non-stationary black will not be, in general, a Killing
horizon). Also, and as in our definition of $\conhoya{C}$, one would
require of any generalization of the entropy formula to reduce to
eq.~(\ref{WaldEntropy}) when evaluated on the horizon of a stationary
black hole. For the interested reader, we review in this Appendix the
solution to this problem found by Iyer and Wald. \\

The basic idea is as follows: instead of modifying the functional form
of the entropy functional $S_{\rm BH}[g,\Sigma]$ given in
eq.~(\ref{WaldEntropy}), Iyer and Wald provided a specific algorithm
to deform the dynamical metric $g_{ab}$ in a neighbourhood of a
spacelike cross-section of the event horizon of a dynamical black
hole. Let us denote such a cross-section by $\Sigma_{\rm dyc}$. This
deformed spacetime, that here we call $\widehat{g}_{ab}$, is such that
$g_{ab} = \widehat{g}_{ab}$ at $\Sigma_{\rm dyc}$. Moreover, it has
the property that $\Sigma_{\rm dyc}$ becomes the bifurcation surface
of a bifurcate Killing horizon of the metric $\widehat{g}_{ab}$.  This
will make the quantity $S_{\rm BH}[\widehat{g},\Sigma_{\rm dyc}]$ to
be automatically well defined (cf. the discussion at the beginning of
Section~\ref{sec:4.1}).  Finally, it turns out that
$\widehat{g}_{ab}=g_{ab}$ when the ``original'' spacetime is that of a
stationary black hole, and therefore $S_{\rm BH}[\widehat{g},\Sigma_{\rm
dyc}]=S_{\rm BH}[g,\Sigma]$. Let us summarize now the prescription of
Iyer and Wald to obtain $\widehat{g}_{ab}$ from $g_{ab}$.

\subsection{Boost-invariant part of $g_{ab}$ at $\Sigma_{\rm dyc}$}
\label{BIPart}

Their algorithm to find $\widehat{g}_{ab}$ is as follows. First
choose, at any point $p$ in $\Sigma_{\rm dyc}$, a couple of
independent null vectors (unique up to scale) $k^a$ and $n^a$
orthogonal to $\Sigma_{\rm dyc}$, obeying the (conventional)
normalization condition $k^a n_a=-1$. Take now a neighbourhood of
$\Sigma_{\rm dyc}$ small enough such that any point $x$ in the
vicinity of $\Sigma_{\rm dyc}$ lies on a unique geodesic orthogonal to
$\Sigma_{\rm dyc}$. Next define the following coordinate system in the
space orthogonal to $\Sigma_{\rm dyc}$\footnote{This coordinate system
was used in~\cite{Kay:1988mu}, and their relation to the properties of
spacetimes with bifurcate Killing horizons was studied there.}: given
$x$, find $p$ in $\Sigma_{\rm dyc}$ and the (unique) geodesic
connecting $p$ and $x$. Parametrize this geodesic such that $x$ is at
unit affine parameter from $p$, and find its tangent vector $v^a$ of
such geodesic at $p$. Finally, assign the coordinates $(U,V)$ to $x$
in the 2-space spanned by $k^a$ and $n^a$ to be the components of
$v^a$ along $k^a$ and $n^a$. Note that, in this coordinate system,
$\Sigma_{\rm dyc}$ lies at $U=V=0$.

The next step is to Taylor-expand every component of $g_{ab}$ in $U$
and $V$ around $U=V=0$ up to some order $N$ (to be fixed below). In
this series expansion, remove all terms which do not contain the same
number of $U$'s and $V$'s. The resulting metric, $\widehat{g}_{ab}^N$,
is called the {\em boost-invariant part of order $N$} of $g_{ab}$ at
$\Sigma_{\rm dyc}$.  Finally, choose the order $N$ of
$\widehat{g}_{ab}^N$ to be equal to the higher derivative of the
metric appearing in the Lagrangian. Note that the boost-invariant part
$\widehat{g}_{ab}^N$ is a function of the single combination~$UV$, and
therefore is of the boost-invariant
form~(\ref{UV_metric})\footnote{Actually, $\widehat{g}_{ab}$ is a
power series of the form~(\ref{GeneralBI}). The difference between
$\widehat{g}_{ab}$ and the boost invariant projection
$\conhoya{g}_{ab}$ defined in the paper is the prescription to fix the
coefficients in~(\ref{GeneralBI}).}. Therefore the surface $U=V=0$
becomes the bifurcation surface of the Killing horizon $UV=0$, the
corresponding Killing vector being $\xi=U\p_U-V\p_V$. All this implies
that the quantity:
\begin{equation}
   \label{Sdyc}
   S_{\rm dyc}[g,\Sigma_{\rm dyc}] \equiv
   S_{\rm BH}[\widehat{g},\Sigma_{\rm dyc}] = -2\pi
   \int_{\Sigma_{\rm dyc}}
   \frac{\partial L(\widehat{g})}{\partial \widehat{R}_{abcd}}\,
   \widehat{\epsilon}_{ab}\, \widehat{\epsilon}_{cd}\,
   \sqrt{h}\, d\Omega\, .
\end{equation}
is automatically well defined and free of all the ambiguities present
in $S_{\rm BH}[g,\Sigma_{\rm
dyc}]$~\cite{Jacobson:1993vj,Iyer:1994ys}. Most importantly, in the
case of a stationary black hole (and therefore $\Sigma_{\rm
dyc}=\Sigma$), it turns out that:
\begin{equation}
  S_{\rm dyc}[g,\Sigma] =
  S_{\rm BH}[g,\Sigma]\, ,
\end{equation}
since, for a stationary black hole, the metric $g_{ab}$ at $\Sigma$ is
automatically boost invariant~\cite{Kay:1988mu,Iyer:1994ys}, which
implies that $g_{ab}$ equals $\widehat{g}_{ab}$ at $\Sigma$. This fact
motivated the proposal of~\cite{Iyer:1994ys} to take $S_{\rm
BH}[\widehat{g}, \Sigma_{\rm dyc}]$ as a possible candidate for the
physical entropy of a dynamical black hole. \\

Notice that the definition~(\ref{Sdyc}) for the entropy of a dynamical
black hole is in principle evaluated on a cross-section of the event
horizon. However, the fact that $\Sigma_{\rm dyc}$ is an event horizon
is not required at any point. This is why we used a very similar
prescription to extend the definition of the entropy
functional~(\ref{WaldEntropy}) to arbitrary spacelike surfaces
(regardless of them being event horizons or not).

\subsection{Comparison between $\conhoya{C}$ and $S_{\rm dyc}$}

Notice that the functional $S_{\rm dyc}[g,{\cal S}]$ has all the
required properties that we discussed in Section~(\ref{sec:4.1}) in
order for it to be well defined on arbitrary spacelike surfaces. So,
in principle, it is a legitimate candidate for a
$\conhoya{C}$-function satisfying the properties~a)-d) established at
the beginning of Section~\ref{sec:ConstructionOfTheConho}. Let us
therefore compare here our proposal for $\conhoya{C}[g,{\cal S}]$ and
$S_{\rm dyc}[g,{\cal S}]$. \\

First, note that it is clear that:
\begin{equation}
  \conhoya{C}[g,\Sigma]=S_{\rm dyc}[g,\Sigma]=
  S_{\rm BH}[g,\Sigma]
\end{equation}
when evaluated on a cross-section $\Sigma$ of the event horizon of a
stationary black hole. However, in general:
\begin{equation}
  \conhoya{C}[g,{\cal S}]\neq S_{\rm dyc}[g,{\cal S}]
\end{equation}
on an arbitrary spacelike surface ${\cal S}$. This is because, since
$\conhoya{g}_{ab}$ and $\widehat{g}_{ab}$ are different metrics, their
associated curvature scalars at ${\cal S}$ will not coincide in
general: only on a black hole horizon it is ensured that (at least to
order $N$ in an expansion of the kind of~(\ref{GeneralBI})) we will
have $\conhoya{g}_{ab}=\widehat{g}_{ab}=g_{ab}$. The reason for
defining the $\conhoya{C}$-function as we did (as opposed to using
Iyer and Wald's boost-invariant part to deform the spacetime metric)
is because, at least in the cases that we have been able to check,
only if $\conhoya{C}$ is defined as $\conhoya{C}[g,{\cal S}]\equiv
S_{\rm BH}[\conhoya{g},{\cal S}]$, the conditions for it to be a
non-decreasing function along outgoing null geodesic flow match the
physical requirements of the theory. \\

Let us explicitly verify this in an example. First note that, since
$\widehat{g}_{ab}$ is also of the boost-invariant
form~(\ref{GeneralBI}), the identity~(\ref{Curvatures}) will also hold
for the curvature invariants associated to $\widehat{g}_{ab}$. This
means that a ``hatted'' analogous of~(\ref{SBHBI}) also holds for
$S_{\rm dyc}[\widehat{g},{\cal S}]$. Considering for definiteness the
case of $AdS$ space, using Iyer and Wald's prescription we had
obtained a $\conhoya{C}$-function given by:
\begin{equation}
  \conhoya{C}\equiv S_{\rm dyc} = 16\pi^2b^2
  (\widehat{L_R}-2\widehat{L_T}\widehat{R}) +
  64\pi^2 \widehat{L_T}\,
\end{equation}
where hats denote evaluation of all expressions in $\widehat{g}_{ab}$
at the surface ${\cal S}$. In this case, the final requirement of
$\conhoya{C}$ being non-decreasing of $r$ would not have been
equivalent to the physical requirements~(\ref{ConsistencyReq}). This
is because the curvature scalars of the boost-invariant part of $AdS$,
when evaluated at a generic surface $r=r_0$, do not coincide with
those of $AdS$ itself. In particular, one finds:
\begin{equation}
  \begin{array}{rcl}
    \widehat{R} & = & \displaystyle
    \frac{9r_0^4 - 4r_0^3 + 10r_0^2 + 1}{6r_0^2(1+r_0)^2}
    R_{AdS}\, , \\[.4cm]

\widehat{T} &=& \displaystyle
    \frac{-12r_0^4+36r_0^3-45r_0^2+8r_0-9}{15r_0^2(1+r_0)^2}
    T_{AdS}\, , \\[.4cm]
  \end{array}
\end{equation}
Therefore, even if we see that $S_{\rm dyc}[g,{\cal S}]$ is well
defined, the conditions for it to be non-decreasing do not coincide
with the physical requirements~(\ref{ConsistencyReq}).  Notice also
that, in general $S_{\rm dyc}[g,{\cal S}]$ has the additional
unsatisfactory property that the conditions that will have to be
satisfied for it to be non-decreasing will depend on the detailed form
of the Lagrangian $L_g(R,T)$.  However, the ghost-free
condition~(\ref{ConsistencyReq}) is generic for any Lagrangian of the
form $L_g(R,P,Q)=L_g(R,T)$.



\end{document}